\documentclass[%
 reprint, 
 superscriptaddress,
 amsmath,amssymb,
 aps,
floatfix,
longbibliography
]{revtex4-1}

\usepackage{graphicx}
\usepackage{bm}

\usepackage{xcolor}
\usepackage[colorlinks=true,citecolor=blue,linkcolor=magenta]{hyperref}

\newcommand{\ben}{\begin{eqnarray}}
\newcommand{\een}{\end{eqnarray}}

\newcommand{\be}{\begin{equation}}
\newcommand{\ee}{\end{equation}}

\usepackage[colorlinks=true,citecolor=blue,linkcolor=magenta]{hyperref}

\setcitestyle{open={[},close={]}}

\makeindex
\begin{document}

\title{A structural hysteresis in the charge density wave transition of 1T-TaS$_2$}

\author{Sharon S. Philip}
\affiliation{Department of Physics, University of Virginia, Charlottesville, VA 22904, USA}
\author{Despina Louca}
\thanks{Corresponding author}
\email{louca@virginia.edu}
\affiliation{Department of Physics, University of Virginia, Charlottesville, VA 22904, USA}

\begin{abstract}
In quasi-two-dimensional 1T-TaS${_2}$, a charge density wave (CDW) prototype, the transition occurs in two steps, from incommensurate (ICDW) and nearly commensurate (NCCDW) and from NCDW to commensurate (CCDW), locked in the resistivity step-wise behavior. The hysteresis observed in the resistivity across the NCDW-to-CCDW transition has roots to a structural hysteresis, where local distortions of the $\sqrt{13}a\cdot\sqrt{13}a$ superstructure are revealed by neutron and X-ray diffraction. The structural hysteresis is due to faulty stars of David (SODs) because of Ta displacements away from the perfect trigonal geometry as well as out of plane S distortions. Furthermore, the superstructure exhibits a 3c$_{o}$ layer stacking order that weakens on warming and fully disappears in the ICDW state. 
\end{abstract}
\maketitle

Charge density waves (CDW) are electronic instabilities that propagate through the lattice and often drive the system to a new crystal periodicity as they spontaneously break the symmetry~\cite{gruner1994density}. The new superlattice modulations introduce a gap in the electron density of states. The mechanism behind such phenomenon, however, has been a subject of ongoing debate, especially in the quasi-two-dimensional (2D) transition metal dichalcogenides (TMDs) 1T-MX${_2}$ ($M = Ti, Ta$ and $X = S, Se, Te$) ~\cite{wilson1974charge,wilson1975charge,choi2017recent,kang2017universal,manzeli20172d,zhu2017misconceptions,chen2017emergence,friend1979periodic}. Several scenarios have been proposed to explain the origin of the gap that include band filling due to layer pairing, electron-phonon coupling or electron-electron correlations leading to a Mott state~\cite{wilson1975charge,fazekas1979electrical,petersen2011clocking, hellmann2010ultrafast, thomson1994scanning, butler2020mottness, wang2020band}. Given their layered nature, the out-of-plane coupling is important to understand the electronic characteristics of these materials, with band structure calculations having shown that opening of a gap at the $\Gamma$ point depends on the orbital order and out-of-plane stacking~\cite{ritschel2015orbital, ritschel2018stacking}. Moreover, density functional theory (DFT) predicts that the metal-insulator transition (MIT) arises not from the in-plane superstructure but from the out-of-plane ordering of the superstructure~\cite{lee2019origin}. To this end, the temperature dependence of the crystal structure of 1T-TaS$_{2}$ is essential, and is investigated by combining single crystal X-ray diffraction with the local structure analysis of neutron powder diffraction.

\begin{figure}
\includegraphics[width=1\linewidth]{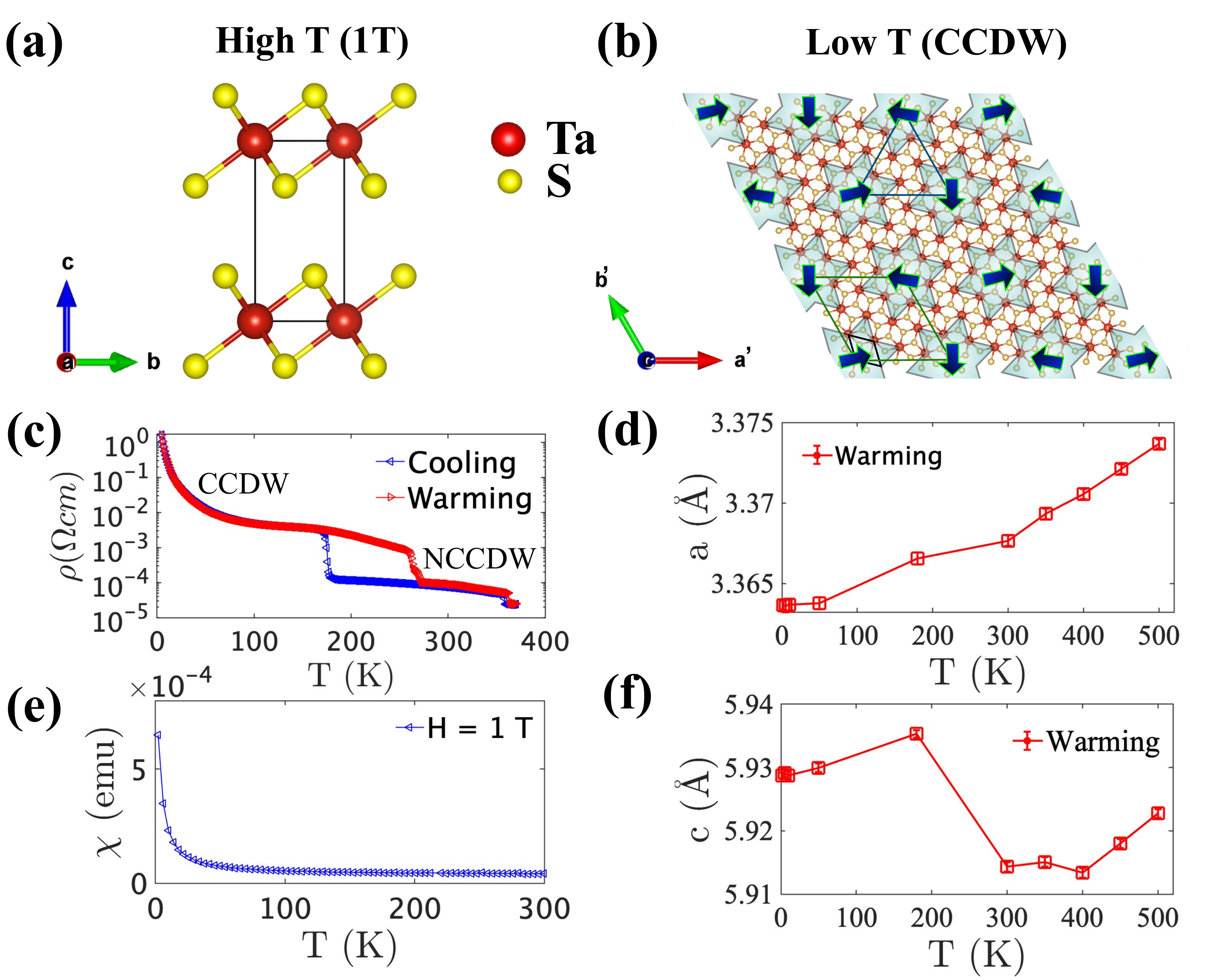}
\caption{(a) The high temperature unit cell with the $P\overline{3}m1$ symmetry of the 1T phase. (b) Schematic diagram of the $\sqrt{13}a\cdot\sqrt{13}a$ supercell in the CCDW phase. Stars of David are colored  green. The uncompensated Ta spin, S = 1/2, is shown by blue arrows at the center of the stars. (c) The resistivity measurements as a function of temperature on a single crystal sample in logarithmic scale. The multiple CDW phase transitions are seen in both cooling and warming. (d) The in-plane lattice constant $a$ as a function of temperature changes slope around the NCCDW-CCDW transition. (e) The susceptibility as a function of temperature at an applied field of 1 T and ambient pressure shows a paramagnetic response at low temperatures. (d) The c-lattice constant showing discontinuity around the CCDW - NCCDW steps and at the NCCDW-ICDW steps.}
\label{Fig_structure}
\end{figure}

\begin{figure}
\includegraphics[width=1\linewidth]{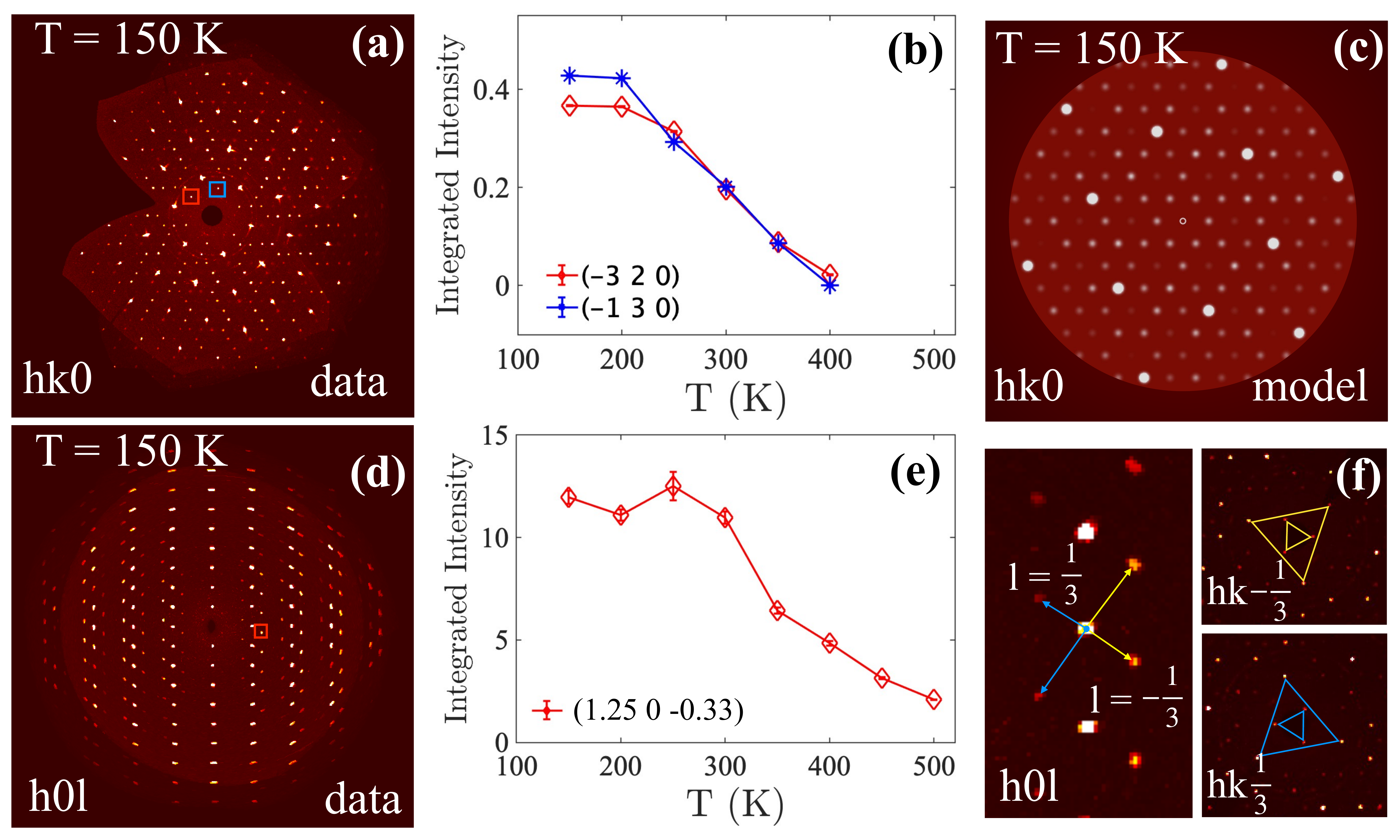}
\caption{Single crystal X-ray diffraction precession image of (a) the (hk0) plane and (d) the (0kl) plane in the reciprocal lattice of 1T-TaS${_2}$ is shown. The weaker supercell reflections, which represents the $\sqrt{13}a\cdot\sqrt{13}a$ lattice, can be seen in between the bright spots. The data were collected at 150 K. (b) The integrated intensity of superlattice peaks in the (hk0) plane as a function of temperature. (c) The diffraction predicted by the $\sqrt{13}a\cdot\sqrt{13}a$ superlattice model is shown in the hk0 plane. (e) The integrated intensity of one of the peaks in (h0l) plane at l=-1/3 as a function of temperature as it undergoes charge density wave transitions. (f) The superlattice reflections along $l=\pm 1/3$ is shown.}
\label{Fig_avgstructure}
\end{figure}

TMD's are fertile ground for coexisting and/or competing non-trivial quantum effects arising from CDW order~\cite{wilson1975charge,chaix2017dispersive}, superconductivity~\cite{dagotto2005complexity,chen2019charge,tranquada1995evidence, tranquada1997coexistence,kivelson2003detect,tranquada2004quantum} and possibly a quantum spin liquid (QSL) ~\cite{balents2010spin,law20171t,anderson1973resonating,anderson1987resonating,han2012fractionalized,zhou2017quantum}. Although a model system to study many-body electron and phonon interactions, TMDs are exploited for engineering  applications~\cite{choi2017recent} and exhibit a multitude of phase transitions and crossovers between proximate states upon cooling, chemical doping, layer stacking, applied strain, or pressure~\cite{choi2017recent,manzeli20172d,ma2018possible,li2021giant,gasparov2002phonon,novoselov2005two,schaibley2016valleytronics,mueller2017tmds,van2016graphene}. Such is the case of 1T-TaS$_2$ with the crystal structure shown in Fig.~\ref{Fig_structure}(a), the host of multiple CDW transitions. Although not superconducting at ambient conditions, superconductivity appears in 1T-TaS${_2}$ under pressure or when doped with selenium (Se) \cite{sipos2008mott,dong2021structural,liu2013superconductivity}.

1T-TaS$_2$ undergoes a unique transformation from the high temperature crystal phase to a commensurate CDW phase that can lead to a quantum triangular antiferromagnet (AFM) (see Fig.~\ref{Fig_structure}(b)). Upon cooling, 1T-TaS$_2$ exhibits three CDW transitions: at very high temperatures, the trigonal structure with $P\overline{3}m1$ symmetry (Fig.~\ref{Fig_structure}(a)) goes through an ICDW phase around $T_{ICDW} \sim$ 540 K due to a Fermi surface instability~\cite{pillo1999remnant}. Further cooling leads to the ICDW becoming NCCDW at T $\sim$ 350 K, at which point the $\sqrt{13}a\cdot\sqrt{13}a$ structural modulation first appears with $\sim$ 12\textdegree \ tilt relative to the ab-plane. Below 180 K, the $\sqrt{13}a\cdot\sqrt{13}a$ structural modulation persists with a rotation of 13.9\textdegree \ relative to the plane where the CDW becomes commensurate (see the commensurate lattice comprised of stars of David in Fig.~\ref{Fig_structure}(b)). The formation of the stars is the result of 12 Ta atoms moving towards a central Ta atom~\cite{wilson1974charge, scruby1975role}. The CDW transitions coincide with the kinks observed in the transport as shown in Fig.~\ref{Fig_structure}(c). The first step occurs at a higher temperature that could not be reached. The second step appears at the onset of the $T_{NCCDW}$, while the third step corresponds to the onset of the $T_{CCDW}$. 

The star of David (SOD) consists of 13 atoms with 12 Ta atoms at the vertices surrounding one lone Ta atom at the center that leaves the band half full. The 12 atom pairs form 6 occupied bands while the 13th atom is left with one unpaired spin in the 5d${^1}$ - d${_z}{^2}$ half-filled band~\cite{fazekas1979electrical,rossnagel2011origin,wang2020band}. The spin-$\frac{1}{2}$ resides on a triangular lattice shown in Fig.~\ref{Fig_structure}(b) that is inherently frustrated and quantum fluctuations prevent magnetic ordering down to the lowest temperature~\cite{law20171t}. The half-filled band is nominally metallic but it is evident from the resistivity that the system becomes insulating. Whether 1T-TaS${_2}$ is a Mott insulator or a band insulator has been highly debated in recent years because the insulating ground state can be explained in terms of out-of-plane stacking without needing to invoke electron correlations. In this work, we show that the in-plane $\sqrt{13}a\cdot\sqrt{13}a$ superstructure is locally distorted due to faults in the SODs that locally break the trigonal symmetry. In addition, out-of-plane sulfur distortions are also present. Moreover, a 3c$_{o}$ layer stacking is observed that exhibits a discontinuous transition below 350 K or so, with the intensity jumping below 300 K in a similar way to the temperature dependence of the c-axis lattice constant as well as the transport upon cooling. The 3-layer stacking order disappears on warming. The results provide evidence for strong electron-lattice coupling.  

Powders and single crystals were prepared using solid state reaction and chemical vapor transport, respectively. Synchrotron X-ray and neutron powder diffraction measurements were performed to investigate the structure through the multiple CDW steps. X-ray powder diffraction measurements were carried out at room temperature using the high-energy beamline (105.7 keV) at the 11-ID-C facility of the Advanced Photon Source of Argonne National Laboratory. The time-of-flight (TOF) neutron diffraction measurements were carried out at the Nanoscale Ordered Materials Diffractometer (NOMAD) at the Spallation Neutron Source (SNS) of Oak Ridge National Laboratory (ORNL) at temperatures ranging from 2 to 480 K. The diffraction data were used for the Rietveld refinement and the pair distribution function (PDF) analysis. The PDF provides a real-space representation of the local atomic correlations without assuming lattice periodicity. Single crystal X-ray measurements were carried out using a Bruker D8 Venture single crystal diffractometer (Mo-$\lambda = 0.71$ \AA) with a Photon III detector and a cryostream that allows for data collection at temperatures between 80 to 500 K. Unit cell refinement and data integration were performed with the Bruker APEX3 software. The transport and bulk magnetic susceptibility results are shown in Figs.~\ref{Fig_structure}(c) and \ref{Fig_structure}(e). A single crystal was used in the transport measurement of Fig.~\ref{Fig_structure}(c) and data were collected from 5 to 370 K. On cooling below 350 K, a jump is observed in the resistivity. This temperature corresponds to the ICDW-NCCDW transition. Upon further cooling, a second jump is observed around 180 K that corresponds to the NCCDW-CCDW transition. The resistivity curve continues to rise as the CCDW state becomes more insulating. The transport behavior in Fig.~\ref{Fig_structure}(c) is consistent with published results~\cite{wilson1975charge, sipos2008mott}. The bulk susceptibility shown in Fig.~\ref{Fig_structure}(e) indicates that this system is paramagnetic down to the lowest temperature.

\begin{figure*}
\centering
\includegraphics[width=1.0\textwidth]{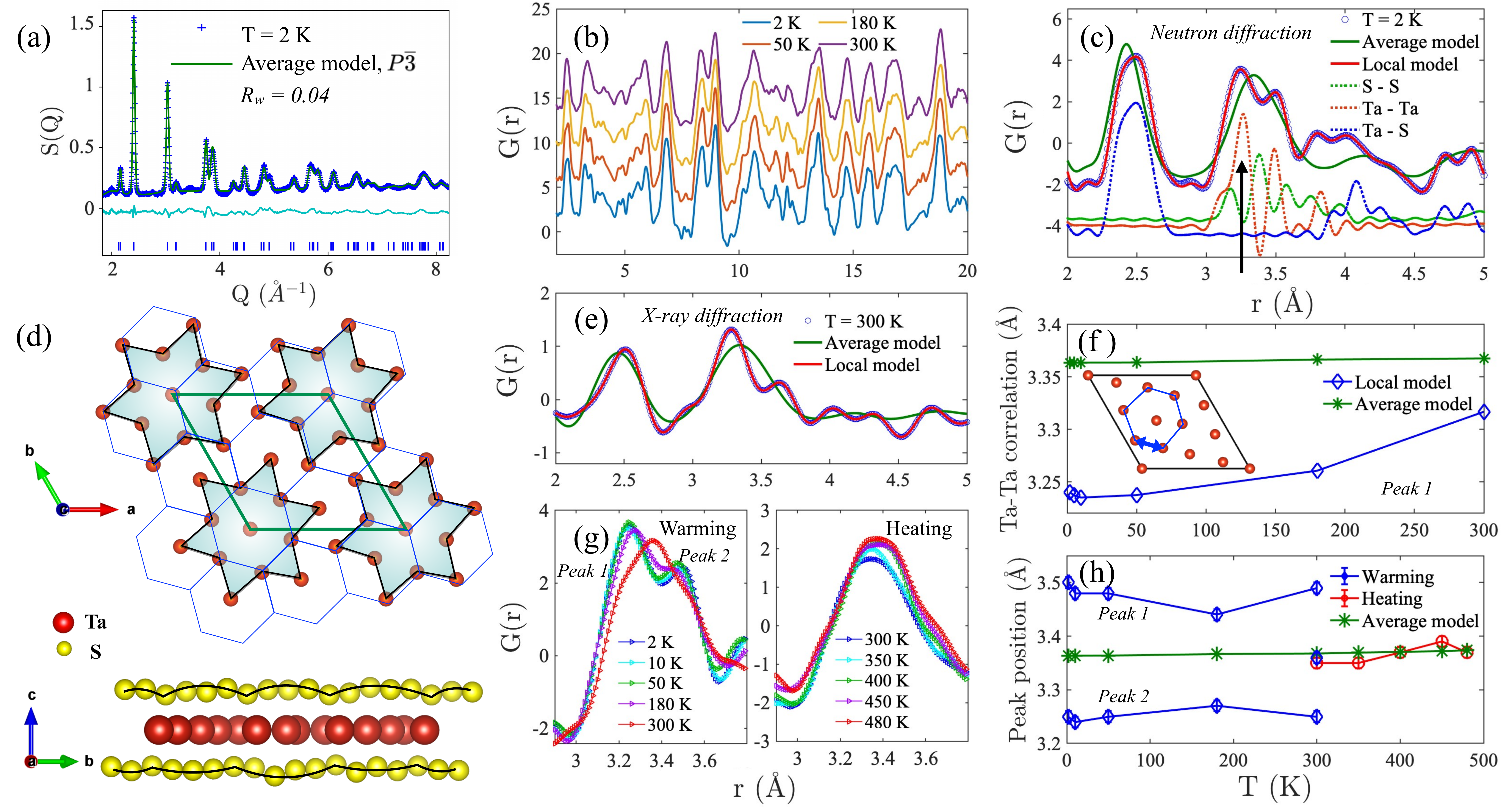}
\caption{(a) Structure function determined from neutron diffraction data at 2 K. The data is compared to the $\sqrt{13}a\cdot\sqrt{13}a$ average model. (b)The experimental neutron PDF as a function of temperature.  (c) The experimental neutron PDF data at 2 K compared with the calculated PDF based on the average model and local model. The partial PDFs describing the individual atom pair correlations is shown as well. (d) The in-plane and out-of-plane local model showing the distorted star of David structures within the CCDW superlattice at 2 K. The blue hexagons are shown to represent the Ta atom positions in the absence of any distortions. The Sulfur atoms (yellow spheres) displace away from the symmetry-restricted planes. (e) The experimental X-ray PDF data at 300 K compared with the calculated PDF based on the average and local model. (f) The Ta-Ta correlation corresponding to peak 1, calculated from the local model and average model. The atom pair is shown in the inset. (g) The evolution of the peak around 3.4 \AA~ is shown from 2 K to 480 K using the data measured upon (left) warming and (right) heating. The peak splitting is labeled as peak 1 and peak 2 corresponding to the peaks on the left and right respectively. (h) The position of peak 1 and peak 2 are shown as a function of temperature upon warming and heating. The predicted position based on the average structure is shown for reference.}
\label{Fig_localstructure}
\end{figure*}

Shown in Figs.~\ref{Fig_structure}(d) and \ref{Fig_structure}(f) are plots of the temperature dependence of the a- and c-lattice constants from neutron diffraction data collected on warming. The observed thermal expansion in either a- or c-lattice constant is not typical, with breaks in the slope observed at the CCDW-NCCDW and NCDW-ICDW transitions as in the transport plot of Fig. 1(c). The discontinuity observed in the a-lattice constant is at or about the CCDW-NCCDW transition (between 200 - 300 K).  Similarly, in the c-lattice constant, a jump is observed between 180 - 300 K, indicating a negative thermal expansion. On further warming, the c-lattice constant remains constant between 300 and 400 K (in the NCCDW phase), after which it rises with temperature above 400 K in the ICDW phase.The lattice constants were obtained from the Rietveld refinement of the diffraction data (Fig.~\ref{Fig_localstructure}(a)). The data were fit using the low temperature symmetry, $P\overline{3}$, that supports the $\sqrt{13}a\cdot\sqrt{13}a$ superlattice. Several superlattice peaks were observed with the reorientation and expansion of the unit cell fronm the high temperature phase. The $R$ factor from the refinement is 0.04. Above 300 K, the $P\overline{3}$m1 symmetry fit the data well. 

In-plane and out-of-plane structural features are mapped to the changes in the resistivity. Precession images collected at 150 K using a single crystal diffractometer are shown in Fig.~\ref{Fig_avgstructure}. At temperatures above the NCCDW-ICDW transition, only the primary Bragg peaks from the $P\overline{3}$m1 are visible. Below this transition, in addition to the primary Bragg peaks, superlattice reflections begin to appear. Fig.~\ref{Fig_avgstructure}(a) is a plot of the hk0-plane in the CCDW phase. The Bragg peaks and superlattice reflections can be reproduced by using the $\sqrt{13}a\cdot\sqrt{13}a$R13.9$^0$ structure. The comparison with the simulation shown in Fig.~\ref{Fig_avgstructure}(c) is exact. The temperature dependence of the integrated intensity of two superlattice reflections is plotted in Fig.~\ref{Fig_avgstructure}(b). The in-plane $\sqrt{13}a\cdot\sqrt{13}a$R13.9$^0$ structure settles below 200 K and the transition to this phase or out of this phase occurs continuously with temperature. By 400 K, the superlattice reflections become indiscernible, indicating that the SODs are either non-existent or quite diffuse that they do not form a pattern. Fig.~\ref{Fig_avgstructure}(d) is a plot of the precession image in the 0kl-plane. Here, we observe additional satellites at every Bragg peak at 1/3 that arise from the type of ordering in the out-of-plane direction. The satellites in the c-direction appear at $l=\pm 1/3$. The temperature dependence of one such peak is shown in Fig.~\ref{Fig_avgstructure}(e). Starting from low temperatures, the 1/3 superlattice intensity is steady until the temperature reaches the hysteresis region of the NCDW-ICDW border, at which point the intensity drops. The drop continues through the ICDW transition temperature and gradually disappears by 480 K. Thus it is clear that the 1/3 satellite exists above 400 K, in the ICDW phase. Shown in Fig.~\ref{Fig_avgstructure}(f) is a plot in the 0kl-plane, providing a closer look at the 1/3 peaks. Cuts at constant $l=\pm 1/3$ as in the smaller figures of Fig.~\ref{Fig_avgstructure}(f) reproduce the Bragg peaks projected on the hk0 plane.

On further probing the structure, we observe a local structural hysteresis that is analogous to the hysteresis observed in the resistivity. The neutron and X-ray powder diffraction data were Fourier transformed to obtain the PDF. Probing the local structure with X-rays and neutrons provided constrains to the real-space model that enabled a detailed picture of the changes emerging with the phase transitions. The G(r)'s as a function of temperature are shown in Fig.~\ref{Fig_localstructure}(b). Changes are observed as a function of temperature that we elaborate on next. Shown in Fig.~\ref{Fig_localstructure}(c) is a plot of the neutron G(r) determined at 2 K and compared to a model calculated using the atomic coordinates and unit cell dimensions of the $\sqrt{13}a\cdot\sqrt{13}a$ superlattice structure. We call this the "average model" as it is based on the symmetry that describes the periodic unit cell. It is evident, however, that that the periodic lattice of the $\sqrt{13}a\cdot\sqrt{13}a$ alone do not adequately describe the local atomic structure(Fig.~\ref{Fig_localstructure} (c)). Differences are observed between the two, and most striking is the discrepancy around 3 to 4.5 \AA. The discrepancy is also evident in the X-ray G(r) data of Fig.~\ref{Fig_localstructure}(e) at 300 K. A better fit to the local structure is obtained by allowing Ta and S atoms displacements leading to the distortions shown in Fig.~\ref{Fig_localstructure}(d). In-plane SOD arrangements are distorted away from the ideal hexagonal geometry, while out of plane S distortions lead to a modulation that runs parallel to the plane. Fig.~\ref{Fig_localstructure}(f) is a plot of the local Ta-Ta correlations at $\sim$ 3.3 \AA~obtained from the fitting of the G(r) data compared to the Ta-Ta bond obtained from the average structure (see inset of Fig. 3(f)). This Ta-Ta correlation is indicated by a black arrow in the partial PDF shown in Fig.~\ref{Fig_localstructure}(c). While the particular Ta-Ta bond does not change with temperature in the average structure, locally it is shorter and changes with temperature as shown in Fig. 3(f).

Also shown in Fig.~\ref{Fig_localstructure}(c) and Fig.~\ref{Fig_localstructure}(e) at 2 K (CCDW) and 300 K (NCCDW), respectively, is a comparison of the calculated G(r) based on the local model of Fig. 3(d). The agreement is very good. In this model, the SODs are faulty in real-space, where the Ta atoms move within the plane and locally break the $P\overline{3}$ symmetry. The Ta distortions are continuous as a function of temperature, with the maximum atom displacements observed at 2 K (Fig. 3(f)). The hexagonal lattice shown in the background of the stars corresponds to the average symmetry of $P\overline{3}$. Thus it is clear that the SOD symmetry is locally broken but such distortions are not sufficient to break the long-range symmetry. Also shown in this figure are the out-of-plane S distortions where a quasi-periodic modulation propagates in the direction tangential to the c-axis. 

Shown in Fig.~\ref{Fig_localstructure}(g) is the temperature dependence of the G(r) in a narrow range between 3 and 3.8 \AA, with the two peaks identified as $Peak~1$ and $Peak~2$. The left panel is for data collected on warming, after the sample was cooled all the way down to 2 K. The right panel shows data collected on heating from 300 K on. From the partial plots of Fig.~\ref{Fig_localstructure}(c), it can be seen that Ta-Ta and S-S pair correlations contribute to the total PDF in this region. The temperature dependence of peaks 1 and 2 are shown in Fig.~\ref{Fig_localstructure}(h). The peak splitting primarily arises from the local Ta-Ta correlations and secondly from S-S correlations. In comparison to the average model where only one broad correlation peak is observed in that region at all temperatures, the real space correlations are split and the split changes with temperature. For data collected on warming, it is clear that the faulty arrangement of the stars is retained through the NCCDW transition. For data collected on heating from 300 K, there is no memory of the faulty stars and only one broad correlation is observed. At 300 K, the data on warming and heating are different because of the structural hysteresis. Thus the resistivity behavior on warming is most likely associated with the structural hysteresis.
Stacking of the CDW ordered planes plays a key role in the formation of the gap at the Fermi level~\cite{ritschel2018stacking}. It has been theoretically proposed that hybridization of the d${_z}{^2}$ orbitals along $c$ can lead to a gap at the $\Gamma$ point. The temperature dependence of the local distortions, formation of the superlattice and out-of-plane stacking order provide evidence for strong electron-lattice coupling that may hold the key to the gap opening in the density of states. Moreover, the band filling scenario is not consistent with our observations because of the presence of distortions. Instead, a more likely scenario must involve electron-phonon and electron-electron correlations.
\\ \\
This work has been supported by the National Science Foundation, Grant number 
2219493. The authors acknowledge Joerg Neuefeind for support provided with the neutron experiment, Yang Ren for assistance provided with the 11-ID-C measurement, and Jianshi Zhou for help with the magnetic susceptibility measurement. A portion of this research used resources at the Spallation Neutron Source, a DOE Office of Science User Facility operated by Oak Ridge National Laboratory. Use of the Advanced Photon Source at Argonne National Laboratory, Office of Science user facility, was supported by the U.S. Department of Energy, Office of Science, Office of Basic Energy Sciences. Single-crystal X-ray diffraction experiments were performed on a diffractometer at the University of Virginia funded by the NSF-MRI program (Grant CHE-2018870).

\bibliographystyle{apsrev4-1}
\bibliography{ref}

\end{document}